\documentclass[a4paper]{JAC2003}
\usepackage{graphicx}
\usepackage{amsmath}
\usepackage{booktabs}
\usepackage{epsfig}
\begin{document}
\title{BEAM-BEAM EFFECTS IN THE SPS PROTON-ANTI PROTON COLLIDER}

\author{K. Cornelis, CERN, Geneva, Switzerland\\}

\maketitle

\begin{abstract}
During the proton-anti proton collider run several experiments were carried out in order to understand the effect of the beam-beam interaction on backgrounds and lifetimes. 
In this talk a selection of these experiments will be presented. 
From these experiments, the importance of relative beam sizes and tune ripple could be demonstrated.

\end{abstract}

\section{GENERAL LAYOUT OF SPS COLLIDER OPERATIONS}

\begin{figure}[h]
\includegraphics*[width=80mm, totalheight=58mm, angle=-00]{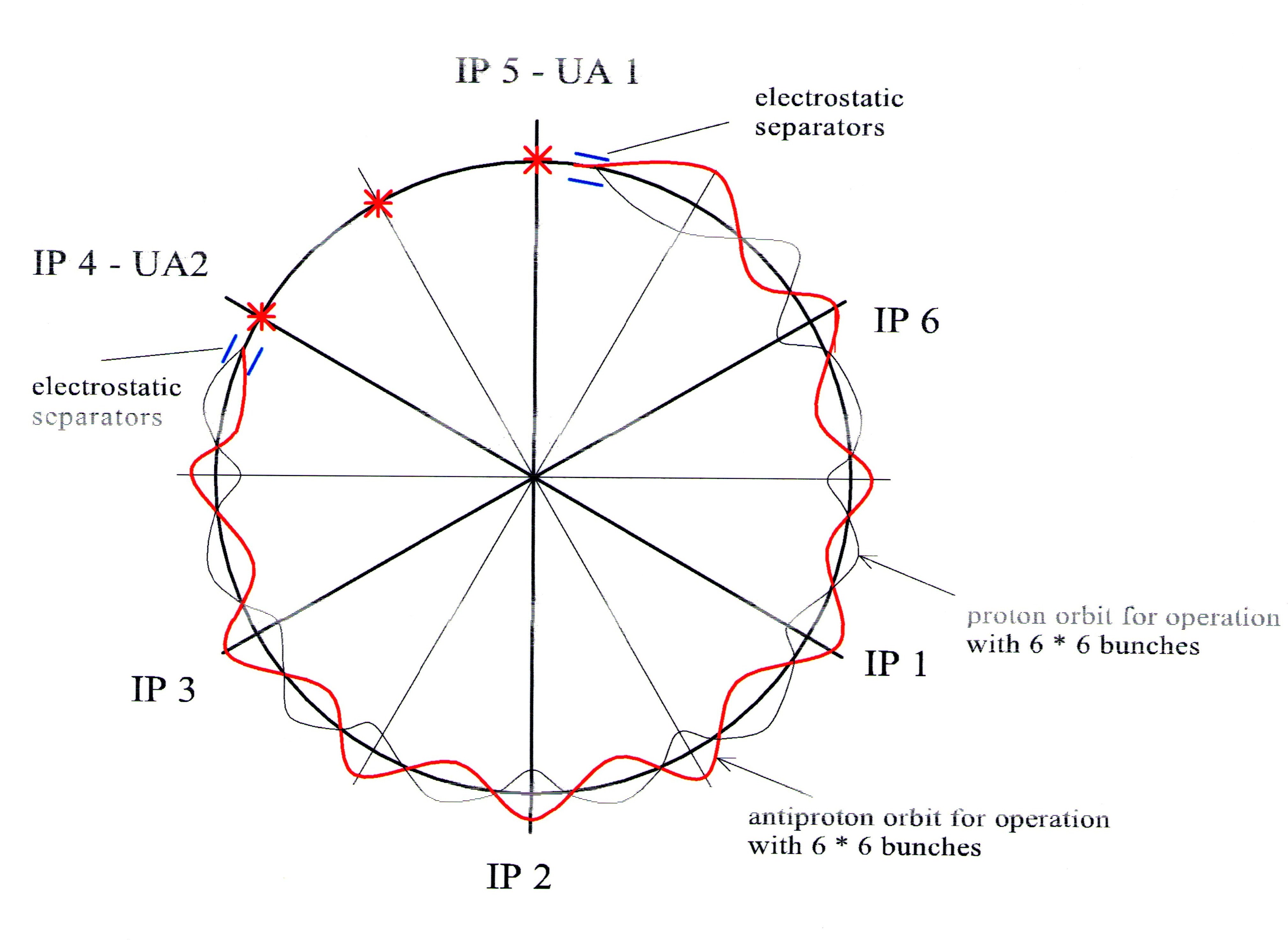}
\caption{Schematic layout of the SPS pretzel scheme.}
\label{fig:pretz}
\end{figure}

In the first collider runs, the Super Proton Synchrotron (SPS) was operated with three proton bunches against three anti-proton bunches, colliding in six collision points. 
The proton bunch intensity at that time was close to $2\times10^{11}$ and the anti-proton intensities were about ten times less. 
In this configuration, total tune shifts of 0.028 were sometimes obtained but the anti-proton lifetimes in the beginning of the coast were poor. 
A horizontal pretzel scheme was introduced to separate the beams in the unwanted collision points (Fig.~\ref{fig:pretz}) and the SPS could then profit from the upgraded anti-proton accumulation facility, operating with six against six bunches. 
In this scheme the beams were separated in 9 of the 12 crossing points.

\begin{figure}[h]
\centering
   \includegraphics*[width=80mm, height=65mm, angle=-00]{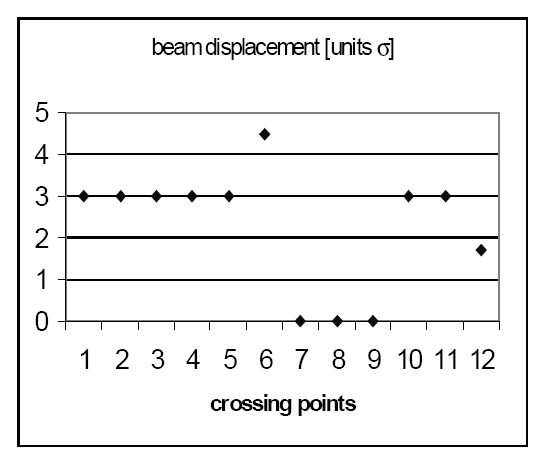}
\caption{Beam half-separation in the 12 crossing points.}
\label{fig:sep}
\end{figure}

The beam separation was 6\,$\sigma$ or better in all the crossing points, except for one, where the separation was only 3.5\,$\sigma$ (Fig.~\ref{fig:sep}).

~~~\\

The same electrostatic separators were used during injection to separate the beams at all crossing points, in order to keep emittances small during the injection and acceleration process.
The parameters of the SPS collider in the final six-on-six operation are listed in Table~\ref{tab:66}.

\begin{table}[h]
	\centering
	\caption{General parameters during the SPS collider run with six protons on six anti-protons. The beam emittance $\epsilon$ is defined as $\epsilon=4\gamma\sigma^2/\beta$. }
	\begin{tabular}{lc}
		\toprule{\textbf{Parameter}}	&{\textbf{value}}		\\
		\midrule
			Beam injection energy	& 26\,GeV				\\
			Beam coast energy		& 315\,GeV				\\
			Proton intensity		& $1.7\times10^{11}$ p/bunch, 6 bunches	\\
			Anti-proton intensity	& $0.8\times10^{11}$ $\overline{\mathrm{p}}$/bunch, 6 bunches	\\
			$\epsilon_x,\epsilon_y$	& 15--20\,mm mrad\\
			$\xi _{x,y}$ 		& 0.015--0.020(total)	\\
		\bottomrule
	\end{tabular}
	\label{tab:66}
\end{table}

\begin{figure}[h]
\centering
   \includegraphics*[width=75mm, height=75mm, angle=-00]{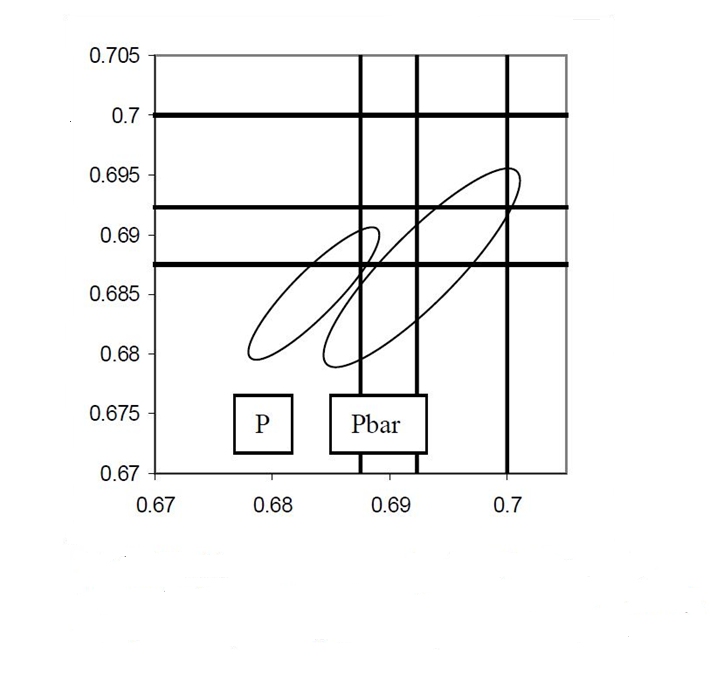}
\caption{Working point during physics runs. The horizontal and vertical lines represent respectively the 16th (.6875), the 13th (.6923) and the 10th (.700) order resonance.}
\label{fig:wd}
\end{figure}

The tune diagram with typical proton and anti-proton footprints during physics runs is shown in Fig.\ref{fig:wd}. 
The tune difference between protons and anti-protons could be trimmed with sextupoles, placed where the beams were separated. 
The core of the beam is at the higher tune values.

\section{THE INFLUENCE OF BEAM SIZE}

\begin{figure}[h]
\centering
   \includegraphics*[width=85mm, height=65mm, angle=-00]{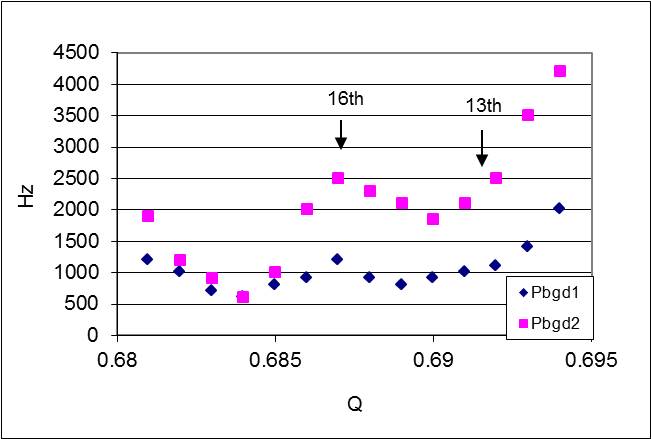}
\caption{Proton background as a function of tune, before (Pbgd1) and after (Pbgd2) the pbar emittance was reduced by 30\%.}
\label{fig:back}
\end{figure}

In some of the runs, where the machine went into coast with substantially smaller pbar emittances, the protons had a substantially shorter lifetime and gave a lot of proton background, and was in spite of the fact that the pbar intensity was 10 times less than the proton intensity. 
This phenomenon could be artificially reproduced by scraping one of the beams with a collimator at a place where they are separated and observing the background and/or the lifetime of the other beam. 
In Fig.\ref{fig:back} a horizontal tune scan is shown looking at the proton background before and after the pbar emittance was reduced by 30\%. 
Although the intensity of the anti-protons was reduced by the scraping process, the protons seem to suffer much more from the beam-beam interaction. 
This experiment is showing that the high amplitude particles (amplitude measured in units of $\sigma$ of the other beam) suffer from the high-order resonances.

\begin{figure}[h]
\centering
   \includegraphics*[width=85mm, height=65mm, angle=-00]{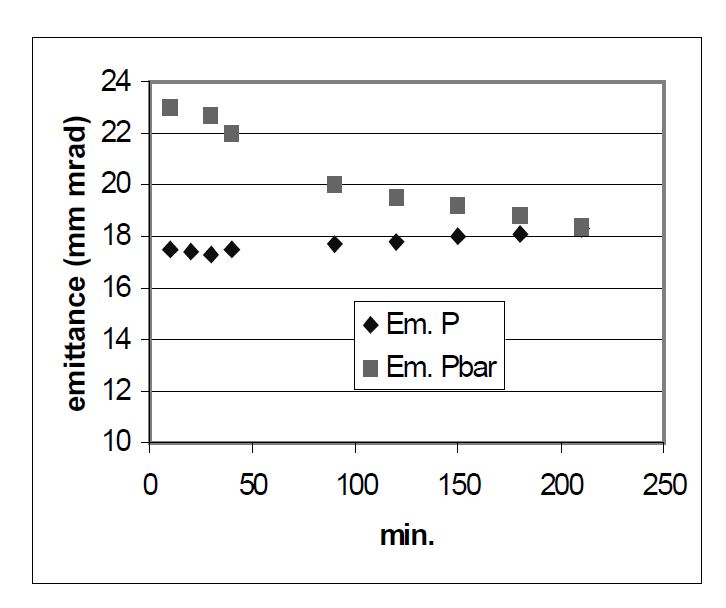}
\caption{Evolution of the proton and anti-proton emittance during the first 200 min.\ of a coast.}
\label{fig:emit}
\end{figure}

In fact it turned out to be very important for the lifetime and background that the two beams had the same size. 
In case of unequal beam sizes, the bigger beam would loose all the particles sitting outside the other beam mainly, creating a low lifetime and high background in the beginning of the coast. 
This is illustrated in Fig.\ref{fig:emit}, where the evolution of the proton and anti-proton emittances is plotted during a normal physics coast. 
By accident the protons are smaller than the anti-protons at the beginning, but throughout the coast the anti-protons mainly lose high amplitude particles and after three to four hours the anti-proton emittance is reduced and matches the proton emittance. 
The small growth of the proton emittance is due to intra-beam scattering.

\section{THE INFLUENCE OF SEPARATION}

\begin{figure}[h]
\centering
   \includegraphics*[width=85mm, height=65mm, angle=-00]{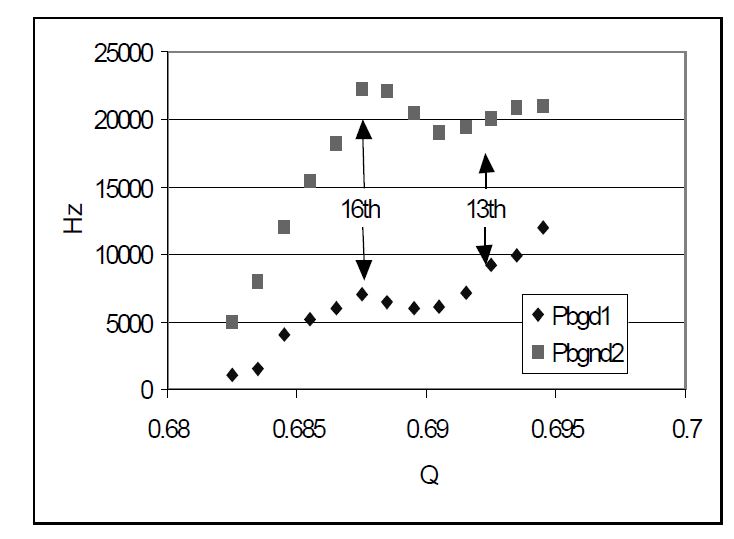}
\caption{Tune scans with full (Pbgd1) and half separation (Pbgd2).}
\label{fig:tunescan}
\end{figure}

\begin{figure}[h]
\centering
   \includegraphics*[width=85mm, height=65mm, angle=-00]{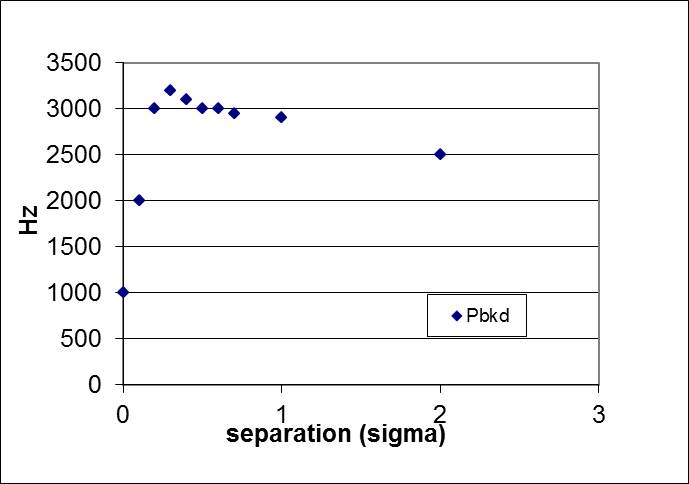}
\caption{Proton background as a function of the separation.}
\label{fig:backsep}
\end{figure}

Tune scans were also performed for different separations.
Reducing the separation in the parasitic crossing from 6\,$\sigma$ to 3\,$\sigma$, increases the background on the 16th as well as on the 13th order resonance as can be seen in Fig.~\ref{fig:tunescan}.
In another experiment only one bunch of protons was colliding head-on with one bunch of anti-protons at two collision points.
The beam was then separated at one of the two points in steps of 0.1\,$\sigma$. 
The result is shown in Fig.~\ref{fig:backsep}. 
The background rises very quickly as a function of separation, reaching a maximum at 0.3\,$\sigma$. 
The background then decreases very slowly as a function of further separation.

\section{THE EFFECT OF TUNE MODULATION}

\begin{figure}[h]
\centering
   \includegraphics*[width=85mm, height=110mm, angle=+00]{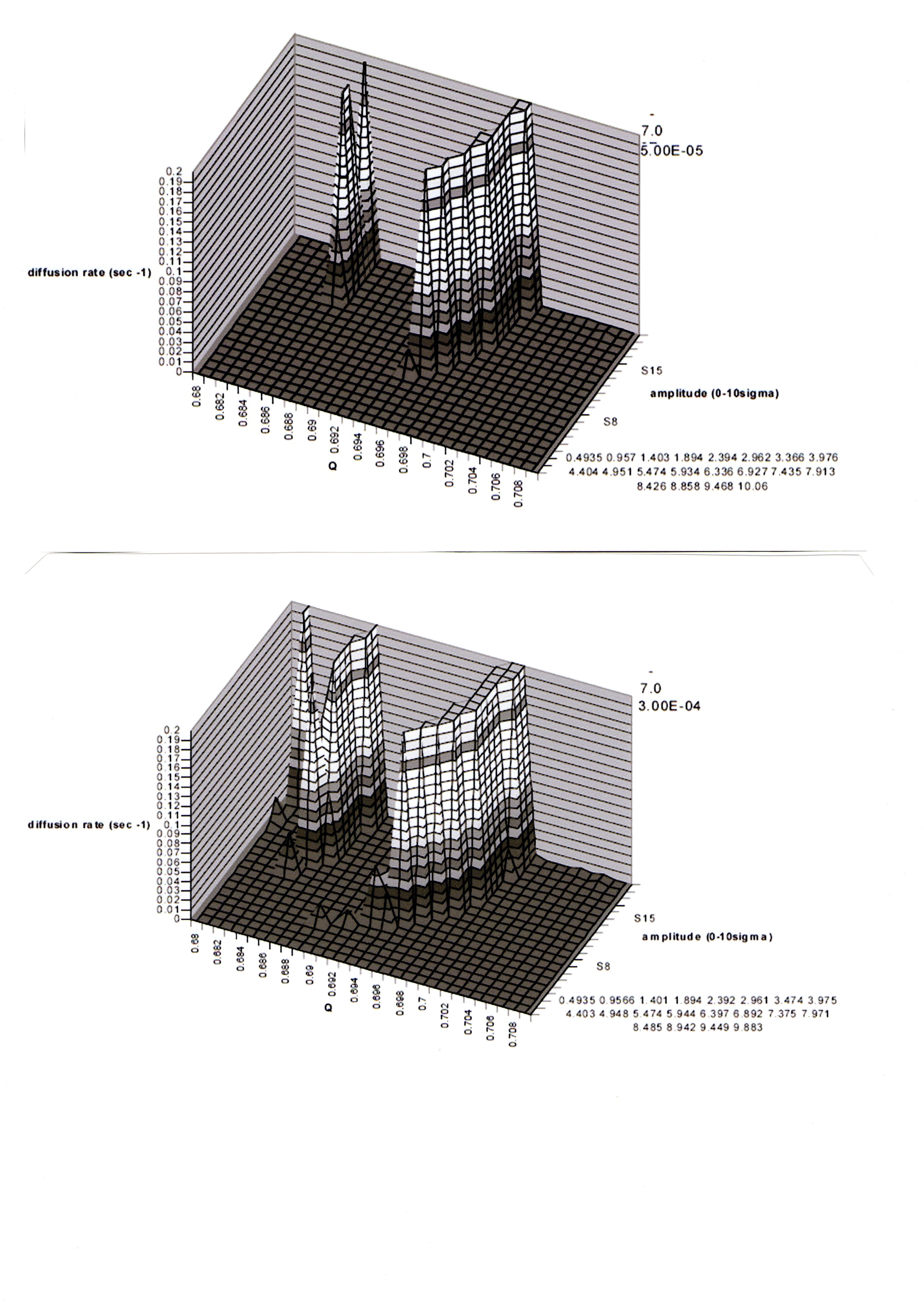}
\caption{Diffusion rates ($z$) as function of tune ($x$) and initial amplitude ($y$: 0 to 10\,$\sigma$). Total tune shift 0.012. Tune modulation 200\,Hz, modulation amplitude 0.000 05 (above) and 0.0003 (below).}
\label{fig:diff}
\end{figure}

High-order resonances in non-linear fields manifest themselves as stable islands in phase space. 
Most of the particles will have a varying amplitude but the motion stays periodic and stable. 
Tune modulation will make islands move inward and outward. 
For small frequencies, the particles will stay trapped in the islands and they can be transported to very high amplitudes due to tune variation. 
If the tune is varied fast enough there will be passages created between the islands through which the
particles can move very quickly inward or outward (cf.\ empty bucket acceleration). 
At even higher frequencies the separatrices of the island become transparent.
Particles can be trapped in an island at a low amplitude and leave the island again at a high amplitude. Whatever the mechanism, tune modulation, together with non-linear resonances, creates the possibilities for particles to move to higher amplitudes.
This could be very clearly observed in the SPS: in order to have good lifetimes with colliding beams, the chromaticity had to be tuned as close as possible to zero.
Also, the noise on the main magnets had be reduced to a minimum in order to preserve good lifetimes during collision.
The effect of tune modulation can be easily demonstrated by simulation. 
In Fig.~\ref{fig:diff} the results are shown of a simple simulation using a head-on beam-beam kick separated by a linear transfer matrix in which the phase advance is modulated with a frequency of 200\,Hz. 
In the first case the tune modulation is $5\times10^{-5}$ and in the second case it is $3\times10^{-4}$. 
The diffusion rate ($z$-axis) is shown as a function of the unperturbed tune ($x$-axis) and the initial amplitude ($y$-axis) ranging from 0 to 10\,$\sigma$. 
In both cases the 16th and 19th order resonance can clearly be seen. 
In the first case, the particles on the 10th order resonance move out with a diffusion time constant of 1 minute or faster only from 4\,$\sigma$ onwards and on the 16th order only from 7\,$\sigma$ onwards. 
In the second case, with the stronger tune modulation, the particles on the 10th order move out with the same speed already at 2.5\,$\sigma$ and on the 16th order already at 5\,$\sigma$.

\section{CONCLUSIONS}

\begin{itemize}

\item Experience with the SPS proton anti-proton collider showed that it is very important to have the same beam sizes for both beams in order to obtain good lifetimes and backgrounds;
\item The high order resonances have almost no effect on the particles with small amplitudes;
\item The 'bad' effect of miss-crossing reaches a maximum at a separation of 0.2 to 0.3\,$\sigma$.
\item All tune modulation (chromaticity, power supply ripple, etc.) should be reduced to a minimum.

\end{itemize}

\end{document}